\documentclass[12pt,preprint]{aastex}
\usepackage{epsfig}

\usepackage{url}
\usepackage{color}

\makeatletter
\def\url@leostyle{%
  \@ifundefined{selectfont}{\def\UrlFont{\sf}}{\def\UrlFont{\small\ttfamily}}}
\makeatother
\newcommand\atel{Astronomer's Telegram }

\date{\today}
\shorttitle{Swift J1539.2$-$6227}
\shortauthors{Krimm et al.}

\begin{document}
\title{Discovery and evolution of the new black hole candidate Swift J1539.2$-$6227 during its 2008 outburst}
\author{ H. A. Krimm\altaffilmark{1,2},  J. A. Tomsick\altaffilmark{3}, C. B. Markwardt\altaffilmark{4},\\  C. Brocksopp\altaffilmark{5},
F. Gris\'{e}\altaffilmark{6}, P. Kaaret\altaffilmark{6}, P. Romano\altaffilmark{7}}

\altaffiltext{1}{CRESST and NASA Goddard Space Flight Center, Greenbelt,
  MD 20771, USA}
\altaffiltext{2}{Universities Space Research Association, 10211
Wincopin Circle, Suite 500, Columbia, MD 21044, USA}
 \altaffiltext{3}{Space Science Laboratory, University of California, 7 Gauss Way, Berkeley, CA 94720-7450, USA}
\altaffiltext{4}{NASA Goddard Space Flight Center, Greenbelt,
  MD 20771, USA}
\altaffiltext{5}{Mullard Space Science Laboratory, University College London, Holmburg St. Mary, Dorking, Surrey RH5 6NT, UK}
\altaffiltext{6}{Department of Physics and Astronomy, University of Iowa, Iowa City, IA  52242, USA}
\altaffiltext{7}{INAF, Istituto di Astrofisica Spaziale e Fisica Cosmica, Via U. La Malfa 153, I-90146 Palermo, Italy}

\begin{abstract}

\noindent We report on the discovery by the {\em Swift} Gamma-Ray Burst Explorer of the
black hole candidate Swift J1539.2$-$6227 and the subsequent course of an outburst beginning in November 2008 and lasting at least seven months.  The source was discovered during normal observations
with the {\em Swift} Burst Alert Telescope (BAT) on 2008 November 25.  An extended observing campaign
with the {\em Rossi X-Ray Timing Explorer} ({\em RXTE}) and {\em Swift} provided near-daily coverage over 176 days, giving us {a good} opportunity to track the evolution of spectral and timing
parameters with fine temporal resolution through a series of spectral 
states.  The source was first detected in a hard state during which strong low-frequency quasi-periodic oscillations (QPOs) were detected.  The QPOs persisted for about 35 days and a signature of the transition from the hard to soft intermediate states was seen in the timing data.  The source entered a short-lived thermal state about 40 days after the start of the outburst.  There were variations in spectral hardness as the source flux declined and returned to a hard state at the end of the outburst. The progression of spectral states and the nature of the timing features provide strong evidence that Swift J1539.2$-$6227 is a candidate black hole in a low-mass X-ray binary system.

\end{abstract}

\keywords{accretion, accretion disks -- black hole physics -- X-rays: binaries}

	\section{Introduction}

\noindent Black hole (BH) binaries are some of the most interesting objects in the galaxy.  Some BH systems such as Cyg X-1, LMC X-1 and LMC X-3 are persistent sources \citep{mcre06} and others such as
GRS 1915+105 \citep{mcre06},  and the recently discovered candidate Swift J1753.5$-$0127 \citep{dura09,zhan07} have outbursts extending for many years. Still other sources such as H 1743$-$322 \citep{prat09,mccl09} and  {XTE J1550$-$564 \citep{aref04} have $\gtrsim 25$\ years between major outbursts, but the systems exhibit fainter rebounds for several years afterwards, while the confirmed BH transient} GX 339$-$4 \citep{kong02} {has} fairly regular outbursts separated by $\sim 1-2$\ y.  However, most BHs are active only occasionally, with intervals between outbursts of many years to decades.  

Although BH binaries exhibit extremely varied behavior, there are some common observational features of such systems (see \citet{mcre06}; \citet[][hereafter \citeauthor*{remc06}]{remc06} for comprehensive reviews).
The first of these is that during outbursts, BH sources pass through different spectral and temporal states, which are indicated by changes in spectral parameters such as power law index and disk temperature and often include large and rapid changes in source brightness.  These states are related to changes in accretion flow and variations in mass accretion rate and possibly to changes in inner disk radius, 
and manifest themselves as differences in what component of the measured X-ray spectrum is dominant. For example, thermal emission from the accretion disk is most significant in the thermal or soft state and power-law emission accompanied by the presence of radio jets dominates in the hard (low-hard) state.  
A second common spectral feature is the presence of a line feature at $\approx 6.5$\ keV, presumably an iron emission line.  A third set of phenomena often seen in X rays is high variability in the emission and the presence of quasi-periodic oscillations (QPOs), both at low frequencies (0.1 to 30 Hz), which are usually variable in frequency and likely related to disk oscillations and magnetic instabilities, and at high frequencies (40 -- 450 Hz) which are more constant in frequency and are probably some sort of resonance phenomenon.  

Many attempts have been made to unify the progression of spectral and timing states and the correlations between observables and to use this information to understand BH accretion processes and derive physical parameters of the black holes.   \citet{fend04} studied the hardness-intensity diagram (HID) and developed a jet-disk coupling model to explain the presence of steady radio jets during the rising hard phase of BH outbursts.  Their model matches tracks in the HID with the radius of the inner disk and the strength and Lorentz factor of radio jets.  They define a ``jet line'' and show that some sources cross and re-cross this line multiple times during an outburst.  \citet{bell10} also discusses the hardness-rms (root mean squared) diagram (HRD) and shows how spectral and timing properties can be used together to define a large number of states and the transitions between them. \cite{smit02} have studied the long-term evolution of spectral hardness with luminosity and show that the simple model of the hard-soft transition as being due to an increase in accretion rate and decrease in the inner disk radius does not cover all observed cases of spectral evolution. 

Other authors such as \cite{yu09} find a correlation between flux at the hard-soft state transition and the peak flux of the soft state and suggest that brightness of the hard-soft transition is strongly influenced by non-stationary accretion and perhaps dependent on the mass in the accretion disk.   Taking a different tack, \citet{shap09} use the correlation between QPO frequency and power law photon index and scale this relationship for different sources to estimate BH mass and distance for sources for which these quantities are not known from other methods.  This is just a sampling of the many avenues of study of BH observables.

When a new BH source is discovered, there is an opportunity to add significantly to our understanding of how such systems work, particularly given the high source to source variation and the low number of confirmed BH binaries: still numbering only a few dozen.  Highly sampled long-term observations can shed great light on the new source, providing information used to improve our general understanding of BH accretion.  We were fortunate to have such extended coverage of the recently discovered BH candidate Swift J1539.2$-$6227 with {\em Swift} \citep{gehr04} and {\em RXTE}.

With its nearly full-sky coverage, the {\em Swift}/BAT \citep{bart05} hard X-ray transient monitor\footnote{http://swift.gsfc.nasa.gov/docs/swift/results/transients/.} has become an important new resource for discovery, particularly of bright, new galactic transients, including pulsars and BH transients. The galactic source Swift J1539.2$-$6227 was first detected with the {\em Swift}/BAT Hard X-ray Transient Monitor on 2008 November 24 (MJD 54794).  {\em RXTE} observations began two days later and continued for a total of 159 observations over the next 176 days, with additional  {\em RXTE} observations two months later and again 15 months later.  

	\section{Discovery and Observations}\label{observations}

\noindent The source Swift J1539.2$-$6227 was discovered in regular {\em Swift}/BAT observations as part of the BAT hard X-ray transient monitor's automated search for previously unknown sources.  It first reached the $6\sigma$\ discovery threshold on MJD 54795, but was found in archived data to have been detected the day before at a lower significance, at $0.024 \pm  0.011\ {\rm counts}\ {\rm s}^{-1} {\rm cm}^{-2}$, and was announced as a new galactic transient on MJD 54796  \citep[2008 Nov 26;][]{atel1855}.  The discovery was confirmed in the {\em RXTE} All-Sky Monitor \citep[ASM;][]{atel1868}. This finding initiated a long series of observations with the {\em RXTE}.  
{\em RXTE} Cycle 12 ToO observations began on 2008 Nov 26 and continued through 2009 March 3 and a further set of Cycle 13 observations continued until 2009 May 21 with additional single observations on 2009 July 26 and 2010 August 26. The median exposure time for the {\em RXTE} observations was 1696~s with 90\% of the exposures between 768~s and 3872~s and 50\% between 1220~s and 2540~s. 
At the time of the discovery, the source was too near the sun for {\em Swift} narrow-field instrument pointing and the first {\em Swift} X-ray Telescope (XRT) observation was carried out on 2008 December 25 (MJD 54825).  A total of six {\em Swift} pointed observations were made.  Table~\ref{tab-xrt} gives the {\em Swift} observing log.

{\em Swift} Ultraviolet/Visible Telescope (UVOT) data were collected at all six epochs.  For five of the six observations, UVOT data were collected with a single filter.  However, during the observation on MJD 54904, multi-filter data were collected.  
In the initial UVOT observations on MJD 54825 and 54826 \citep{atel1893} a bright source was found in a 3906-s co-added uvw2\footnote{The UVOT filter bandpasses are defined in \citet{romi05}.} exposure at the location:
\begin{eqnarray}
RA (J2000) = 15:39:11.963\ \ (234.79985^{\circ}) \nonumber \\
Dec (J2000) = -62:28:02.30\ \ (-62.46731^{\circ})\nonumber \\
G_{lon} = 321.018595^{\circ}	\ \ \  G_{lat} = -5.642750^{\circ}\nonumber 
\end{eqnarray}
with a positional uncertainty of $0.5''$ (90\% confidence). There are no known sources coincident with this position found in the USNO-B1, USNO-A2 or 2MASS catalogs.   At the time of the observation the source magnitude was $18.07 \pm 0.03$\ (uvw2) and $17.96 \pm 0.04$\ (uvm2).
 
\citet{atel1958} reported optical spectroscopy from Swift J1539.2$-$6227. A single 750-s spectrum of the optical counterpart to the X-ray transient was acquired with the MagE echellette spectrograph on the Magellan-Clay telescope at Las Campanas Observatory on 2009 Feb 23 09:28 UT. 
Besides a blue continuum, no Balmer lines (in emission or absorption) were detected, nor was there evidence for He II 4686~\AA\ or Bowen blend emission.
The non-detection of emission lines, which are often found in the optical spectra of X-ray transients, is rare but not unprecedented. The lack of any emission features in the outburst spectrum, combined with the faintness of the source in quiescence, suggests a low-mass main sequence or degenerate donor star companion to the compact accretor.

There are no other known optical observations of this source.  Two radio observations were carried out at the Australia Telescope Compact Array but preliminary analysis suggests that the source remained undetected (Corbel, private communication). 

	\section{Data Reduction and Analysis}\label{lightcurves}

\subsection{RXTE data reduction}

\subsubsection{PCA}\label{PCA-reduc}

The {\em RXTE} Proportional Counter Array \citep[PCA;][]{2006ApJS..163..401J} consists 
of five individual Proportional Counter Units (PCUs), each with 
geometric collecting a<rea of $1500\ {\rm cm}^2$, with effective sensitivity in 
the energy band 2 -- 60 keV.  PCA can observe in multiple data modes simultaneously.  In this work we use Standard2, GoodXenon and Event data modes.  The Standard2 mode provides modest spectral 
resolution with 16-s duration accumulations, and was used for energy 
spectral analysis.  The GoodXenon and Event modes provide event-by-event X-ray 
data useful for high time resolution studies ($1 \mu s$\ resolution for GoodXenon and $125 \mu s$\ for Event mode).  Even the coarser time resolution is more than adequate to resolve the low frequency timing features observed for  Swift~J1539.2--6227.  During these 
observations, each of the PCU on/off states was operated independently 
in order to manage detector breakdown.  We filtered data by removing 
times of earth occultation and passage of the spacecraft through the South Atlantic Anomaly.

For energy spectral analysis, the Standard2 data were accumulated for 
enabled PCUs, excluding PCU0 and PCU1 detectors.  These detectors were 
excluded because they no longer have a propane-layer veto capability.  All PCA layers were analyzed.
The accumulation interval varied, with each spectrum corresponding 
approximately to one RXTE orbit or less.  The estimated background was 
subtracted using the VLE model (dated 2005-11-28), computed by 
 {\ttfamily pcabackest} version 3.7.  The systematic errors of the VLE model are approximately 2\% of the 
background level \citep[1$\sigma$;][]{2006ApJS..163..401J}. Different systematic problems can arise from the use of either the VLE or faint background models; such problems can become important when the source is faint. 
To study this we attempted to use the L7 ``faint'' model as 
well, but the results did not change significantly, even at low flux 
levels.  We therefore use the VLE background model throughout this 
paper. 
In addition, as we describe in Section~\ref{spect-anal}, we found a residual count rate even when the source is not active.  We used a PCA observation one year after the end of the outburst to measure and model this emission, and subtract it from the data during outburst.
The PCA response matrix was estimated using 
 {\ttfamily pcarmf} version 11.6.
 
\subsubsection{HEXTE}

We extracted source and background energy spectra from the data from  the {\em RXTE}
High Energy X-ray Timing Experiment \citep[HEXTE;][]{rothschild98}.
HEXTE alternately points (or ``rocks'') on and off source to obtain
source and background spectra.  Although there are two ``clusters''
of HEXTE detectors, at the time that the Swift~J1539.2--6227
observations were made, the rocking mechanism was only operating
for the ``B'' cluster, and we used only data from this cluster for the
spectra in this work.  We processed the HEXTE event list data
with tools from the HEASOFT software package.  To extract the
spectra, we used the {\ttfamily seextrct} program to produce
spectra with 256 energy bins.  Then, we applied the program
{\ttfamily hxtdead} to the source and background spectra separately
in order to correct the total exposure time for deadtime.  We used
{\ttfamily hxtrsp} to produce the appropriate response matrix.
Finally, we rebinned the HEXTE spectra from 256 energy bins to
11 energy bins in the range 17-240 keV, and we used these spectra for
the spectral fits described below.

\subsection{{\em Swift} data reduction}

\subsubsection{XRT}

The {\em Swift} telescope obtained pointed observations of Swift J1539.2$-$6227 on six occasions, between 2008 December 25 and 2009 May 15 (see Table~\ref{tab-xrt}). 
{\em Swift}/XRT \citep{2005SSRv..120..165B} observations were carried out in Photon-Counting (PC) mode for the first three and in Windowed-Timing (WT) mode for subsequent ones. The PC mode retains full imaging and spectroscopic resolution, but the time resolution is only 2.5 seconds. The instrument is typically operated in this mode only at very low fluxes (useful below 1 mCrab). For Swift J1539.2$-$6227 observations, PC mode was used when the flux of the source was quite high ($\sim 0.1\ \mathrm{Crab}$) which implies some severe pile-up. WT provides time resolution of 1.7 ms and one-dimensional spectral information which is suited for sources with fluxes between 1 and 600 mCrab. The source count rate was below 100~mCrab when this mode was used so that there is no pile-up problem \citep{2006A&A...456..917R}.

Using the {\ttfamily xrtpipeline} routine, we re-extracted cleaned event files from level 1a products using all of the default screening parameters with the exception of the bias correction option. 
This allows us to correct the value of the bias for each CCD frame if the difference between the bias value acquired during the slew to source and the one coming from the last 20 pixels of each CCD frame (and averaged over the last 20 frames) is higher than a threshold. However the correction is not perfect. Specifically, the four individual spectra coming from the 2009 April 11 observation show different energy offsets (even after bias correction) which are prominent at energies below 1 keV, leaving some features at 0.5 keV. Thus we excluded the 0.3-0.6 keV range from the subsequent spectral analysis.
The remaining steps to extract the spectra were quite straightforward and follow the {\em Swift}/XRT Data Reduction Guide \citep{Capalbi2005}.
For the WT data, we extracted regions of interest (i.e. containing the source) using detector coordinates and with a box shape of 40 pixels in width, corresponding to 90\% of the point-spread function at 1.5 keV. Background regions were extracted in the same way by using one region located either left or right of the source. The extraction box is sometimes slightly smaller than 40 pixels to avoid bad detector columns.  
To deal with the pile-up issue in the PC spectra, we extracted the spectra using an annulus with increasing inner radius to exclude the central piled-up part of the source \citep{2006ApJ...638..920V}. We used an inner radius of 16 pixels ($\approx 38 \arcsec$) for the two 2008 December observations and 10 pixels ($\approx 24 \arcsec$) for the 2009 March 8 observation. Outer radii were respectively 30 and 25 pixels.
The background was taken as a 40 pixels circular {(not annular)} region {well} away from the source. 

We then produced Ancillary Response Files (ARFs) for each spectrum using the HEASOFT 6.9 tool {\ttfamily xrtmkarf}. Exposure maps were calculated using {\ttfamily xrtexpomap} and included in the ARFs to account for bad pixels and columns in the CCD. The Response Matrix File (RMF) was taken from the calibration database (CALDB 20091130). 
Individual spectra belonging to the same observation were combined with the {\small FTOOLS} {\ttfamily addspec} routine to increase the signal to noise ratio. This tool permitted us to create PHA spectra associated with combined background and response files.
Finally data were grouped to have at least 20 counts per energy channel through the use of the task {\ttfamily grppha}.

\subsubsection{BAT}

The discovery of Swift J1539.2$-$6277 was made using an extension of the {\em Swift}/BAT hard X-ray transient monitor developed to search for new sources.  The sky images for each individual {\em Swift} pointing, which are used to derive the monitor light curves for known sources, are combined to produce six mosaic sky maps covering the entire sky.  These mosaic maps cover time scales ranging from one to 16 days.  As part of the monitor pipeline, each mosaic is searched for any excess above $5\sigma$\ not matching a catalog source.  A somewhat higher threshold of $6\sigma$, either in two consecutive one-day mosaics or one multi-day mosaic is used to trigger a new source announcement.  The mosaic maps are also scanned to search for any past activity from the new source.

Although ultimately the BAT spectra were not used in the analysis, spectra for the pointed observations were produced and checked for consistency with the other fits.  These observations were processed using the HEASOFT {\ttfamily  batsurvey} script to produce 
eight-channel light curves which were then converted to spectra covering the 
energy range 14-195 keV.  We used only data from Swift pointings in which Swift~J1539.2$-$6227  was in the center of the BAT field-of-view, so there was no need for an 
off-axis correction.  

\subsection{X-ray Spectral analysis}\label{spect-anal}

\noindent We were able to fit the combined PCA and HEXTE data for 150 of the 160 observations.  The statistics of the remaining spectra were too poor to allow meaningful fits. For most of the outburst, the BAT rate was low and inclusion of the BAT spectra did not improve constraints on the fit parameters (although when tested, BAT results were consistent with those of the other instruments).  Therefore, for consistency we did not use the BAT spectra for any of the fits.  Since XRT spectra were only collected at six discrete times, they are used for determining the absorption column (see below) and for cross-calibration.  

Before finalizing our choice of fit model we tried various model combinations for three epochs at three different stages of the outburst (MJD 54799 early in the outburst, 54815 near the peak of the outburst and 54864 during the decay phase, but while the source was still bright enough that the statistical quality of the spectra would be good).  All of the trial model sets included an absorbed power law.  It was clear also that a thermal model was needed and that the fits were improved with the addition of an iron line feature.   To account for reflection from the accretion disk, we tried a variety of {\ttfamily xspec12}\footnote{http://heasarc.gsfc.nasa.gov/docs/xanadu/xspec/} models -- {\em pexriv}, an exponentially cut off power law spectrum reflected from ionized material, {\em reflion}, reflection by a constant density illuminated atmosphere, and {\em kdblur},  a convolution model used to smooth a spectrum by incorporating relativistic effects from an accretion disk around a rotating black hole \citep{laor91}.  We also tried modeling a smeared edge with {\em smedge}.   Overall, the best results were derived from the combination (using xspec12 models): {\em constant * (wabs * (diskbb + cutoffpl+ kdblur * gaussian))}.   The Gaussian line width was set to zero and the line was smoothed by the relativistic effects of the {\em kdblur} model.   All quoted errors are at the 90\% confidence level.

We then made joint preliminary fits with PCA, HEXTE and XRT and found consistent results for the first four XRT observations.  However, for the last two observations, discrepancies between PCA and XRT fits suggested that the PCA spectrum was contaminated by background at a level which became significant at a low source flux ($\lesssim 5\ {\rm counts}\ {\rm s}^{-1}$).   The origin of this background is not immediately clear, but we suspect galactic ridge emission, which 
could be diffuse or a combination of weak point sources in the field-of-view.  Swift J1539.2$-$6227 is at   $G_{lat}\ =\ -5.6^{\circ}$, which is, however, rather far from the galactic ridge \citep{vali98} and from the galactic center ($G_{lon}\ =\ 321^{\circ}$).  There are also no bright ROSAT sources nearby and no reason to expect that this background component would vary with time.
To independently measure the residual background we were awarded one further {\em RXTE} observation more than a year after the end of the outburst when the BAT and ASM light curves showed no continued source activity, 2010 August 26 (MJD 55434) for 2656~s.  The count rate (3-25 keV) for this observation was $2.61\ {\rm counts}\ {\rm s}^{-1}$, consistent with the rate seen late in the outburst.  We subtracted this background rate from the PCA light curve and we fit a simple spectral model to the background data.  We first used an absorbed thermal bremsstrahlung model plus a Gaussian line (xspec {\em wabs*(bremss + gaussian})), but we found that inclusion of the Gaussian component did not improve the fit, so that component was not included in the final model.  Likewise,  diffuse gas emission models ({\em mekal; raymond}) did not improve the fit, so we opted for the simplest model.  The background fit parameters are $N_H\ =\   (1.81 \pm  2.1) \times  10^{22}$ atoms/${\rm cm}^2$;  $kT\  =\ 5.5   \pm 1.7$\ keV;   norm =  $0.0052  \pm  0.0023$.
This model was included in all of the fits as a fixed component.  This method is similar to that used by \citet{cori09} for GX 339$-$4.

In order to ensure consistency and to facilitate the processing of a large number of spectra, we developed an automated script which fit each spectrum with the same models using results from the previous spectrum in time sequence as initial parameters.  
{Two of the model parameters in the  {\em kdblur} model, emissivity index and inclination, were explored in an initial run and
frozen at their average values in the final run
(shown in Table~\ref{tab-spect}).  
The outer accretion disk radius was chosen at a large value typical for such systems, 
leaving only the inner disk radius as a free parameter for {\em kdblur}.} 
The value for $N_H$\  ({\em wabs}) was taken from the average of the fits to the joint PCA/HEXTE/XRT spectra.   

\subsection{Timing analysis}

\noindent  We analyzed all segments of PCA data to search for timing features. 
We generated power spectra using Fast Fourier 
Transforms (FFTs) from data taken in GoodXenon mode.  GoodXenon event 
data in the 2.1-- 33 keV band and the frequency range of $8.0\ \times\ 10^{-3}$\ Hz to 2048 Hz were binned into a light curve with a 4.096-kHz sample rate, and divided into 64-s time intervals.  Light curve 
segments were transformed into power density spectra by FFT, and the 
individual segment spectra were summed over individual RXTE orbits. 
These spectra were searched for both pulsations and quasi-periodic oscillations (QPOs).  No pulsations were found in any observation, with a 2$\sigma$\ limit from the brightest observation of {0.15\%} 
rms amplitude.  Also no high frequency QPOs were seen, but in about a quarter of the spectra we saw clear signals of low-frequency QPOs, at frequencies ranging from 0.17~Hz to $\approx7.5$~Hz.  

For variability analysis, we rebinned the power spectra logarithmically to  a bin width/frequency ratio of 0.03.
The rebinned power spectra were fitted 
with a broken power law continuum model with the addition of a Gaussian 
QPO.  The parameters of the model are the break frequency, the power 
index above and below the break frequency, and the centroid, 1-$\sigma$ 
width and amplitude of the QPO.  The coherence parameter ($Q\ =\ \nu/\Delta\nu$) ranged from 5 to 20 and the rms amplitude as a fraction of the mean count rate, $r$,  was in the range of 1 -- 15\%. The power continuum was fit for 119 power spectra and $r$\ ranged as high as 30\%.  The timing properties at different stages of the outburst are discussed in more detail in Section~\ref{results-timing}.
  
\section{Results}\label{results}

\noindent The brightness of the outburst of Swift J1539.2$-$6227 and our long series of observations allow a detailed study of the source as it passes through a series of states.  We are able to track the evolution of the source in intensity (Section~\ref{results-lcs}), spectral fit parameters (Section~\ref{results-spec}) and timing parameters (Section~\ref{results-timing}).

\subsection{Light Curves}\label{results-lcs}

\begin{figure}
\plotone{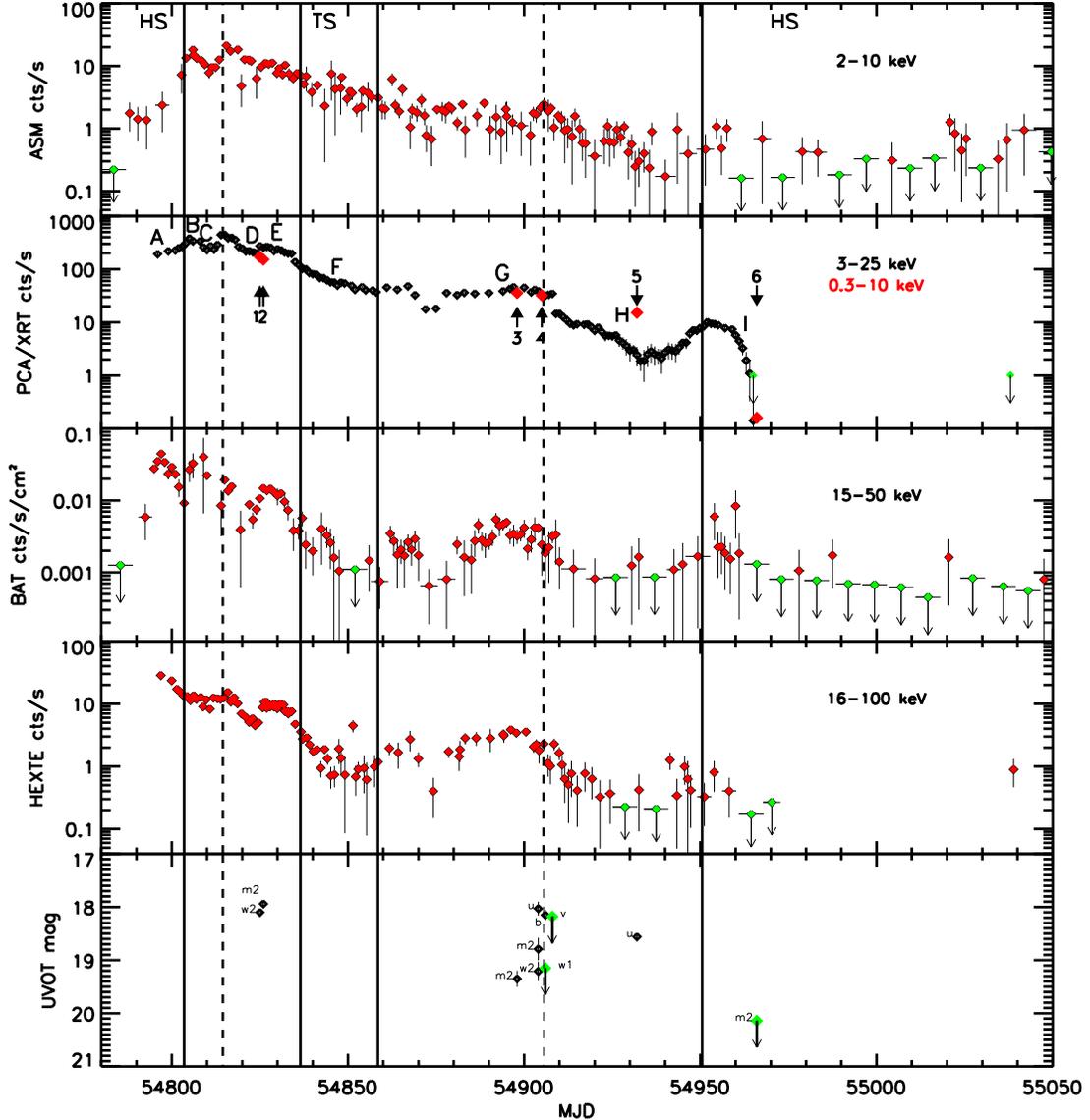}  
 \caption{Light curves for Swift J1539.2$-$6227 for the six instruments,  {\em RXTE}/ASM, {\em RXTE}/PCA, {\em Swift}/XRT , {\em Swift}/BAT, {\em RXTE}/HEXTE and {\em Swift}/UVOT.   On the PCA plot (from which the residual background has been subtracted), 
the times of the {\em Swift}/XRT observations are indicated by arrows.  The solid vertical lines in this and subsequent plots represent the end of the early and start of the late hard states, and the duration of the thermal state as discussed in the text (Section~\ref{discussion-states}).  The dashed vertical lines indicate the transitions from the hard to soft intermediate states. On the BAT, ASM and HEXTE light curves red (diamond) points are detections (error bars do not include zero) and green (crosses in black and white version) are upper limits.  In the second panel, red points  (crosses in black and white version) are XRT ${\rm counts}\ {\rm s}^{-1}$. The letters in the PCA panel correspond to the turning points in the hardness-intensity plot in Figure~\ref{hr_fig}. The actual energy range for the PCA data (approximated in the plot labels) is 2.87 to 25.11 keV.}\label{lin_lc_fig}
\end{figure}

\noindent The light curves for four instruments are shown in  Figure~\ref{lin_lc_fig}.   The BAT and ASM light curves are taken from the public web pages.  The PCA and HEXTE light curves are based on the current work as are the points for  the XRT and UVOT.  Note that the PCA light curves are corrected  {for two different types of background.  The first type is variable in time and is estimated using the  {\ttfamily pcabackest} tool.  The other type is the residual background observed on  2010 August 26 (see Section~\ref{spect-anal}), which we assume to be due to diffuse emission and thus constant.}
The two survey instruments BAT and ASM (from which source rates or limits can be derived from before the first detection) show that the rise to the outburst is very rapid.  In the BAT the flux rose by a factor of nearly 100 over a few days. This is comparable to the rapid rises seen for other sources in outburst such as GRO~J1655$-$40 \citep{harm95} and XTE~J1550$-$564 \citep{hann01}. The rise in the ASM is less sharp, but it is clear that the onset to the outburst is much steeper than the light curve decay, which is common for X-ray transient events. After the rise,  all of the light curves show the same general pattern of a gradual decline punctuated by rebrightenings lasting $\approx 5\ -\ 10$\ days.  Several of the sharp light curve features also match spectral and timing features, particularly in the PCA (see Section~\ref{discussion-states}).  
The general rate of decay in the PCA is an e-folding decay time of $\approx 30$\ days.  By the end of the main observing campaign at MJD 54972, the source was no longer visible in the BAT or HEXTE, and was only slightly above background in the ASM.  A single late {\em RXTE} observation at MJD 55038 showed a level in the PCA consistent with background (and subtracted out in Figure~\ref{lin_lc_fig}) and a positive, though low significance, flux in the other instruments, probably attributable as well to a diffuse or unresolved background component.

The BAT data before the outburst were searched at a one-day time scale back to 2005 February for any previous activity from the source.   No activity was found with a $3\sigma$\ limiting magnitude of 20~mCrab.  There is also no detectable activity to $\approx 50$\ mCrab in the ASM going back 12.9 years before the first detection.

While quite sparsely sampled, the UVOT light curve (bottom panel of Figure~\ref{lin_lc_fig}) shows a decline (seen most clearly in the m2 filter) of approximately 1.1 magnitudes between observations separated by $\approx 70$\ days, which is roughly consistent (to within a factor of 2) with the decline seen in the X-ray rates.  

\subsection{Spectroscopy}\label{results-spec}

\begin{figure}
\epsscale{0.9}
\plotone{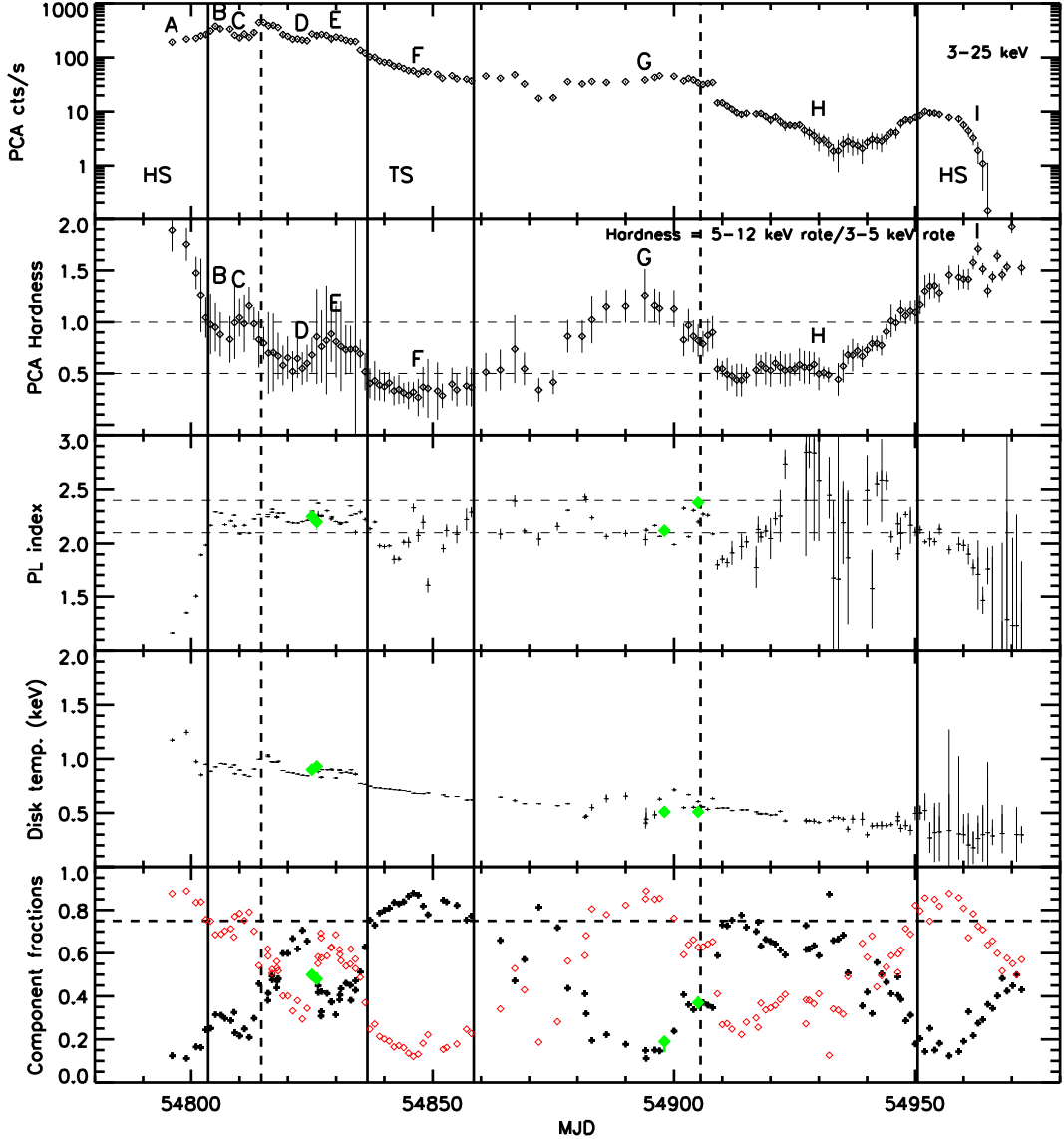} 
  \caption{Hardness ratio and fit parameters.  The top panel repeats the PCA light curve from Figure~\ref{lin_lc_fig}.  The second panel shows the PCA hardness ratio.  The energy ranges are 4.81--11.94 keV/2.87--4.81 keV (approximated in the plot labels). The letters correspond to the turning points in the hardness-intensity plot in Figure~\ref{hr_fig}.  The third panel shows the power-law index. The fourth panel shows the disk temperature for a black body model.  The bottom panel shows the fraction of the flux in each of the primary models --  Red (crosses in black and white version): power law, black dots: disk.  The vertical lines are defined in the caption to Figure~\ref{lin_lc_fig}. The horizontal lines are defined as follows. In the second and third panels, lines show the boundaries of the hard (above top line) and soft (below bottom line) states, as in \citet[][for second panel]{mcre06} and \citeauthor*[][for third panel]{remc06}.  The line in the bottom panel indicates 75\% component fraction.  On the three bottom plots the parameter values from the joint XRT/PCA/HEXTE fits are shown as green (open in black and white) diamonds, indicating consistency of the fits.}\label{fit_pars_fig}
\end{figure}

\begin{figure}
\plotone{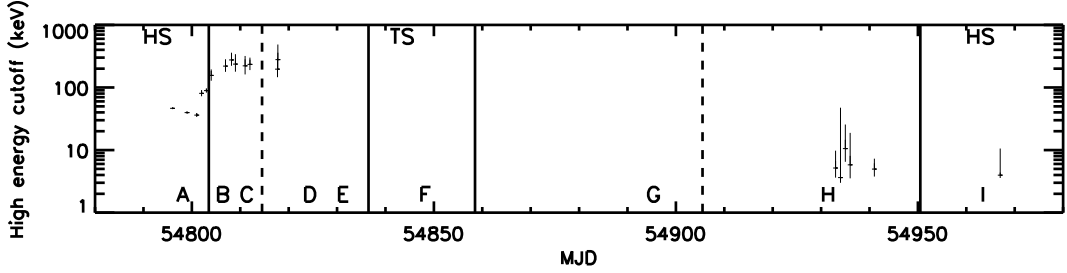} 
  \caption{Fits to the high energy cutoff energy are plotted when this parameter is constrained.  The vertical lines and letters are as defined in the caption to Figure~\ref{lin_lc_fig}. The data range does not allow a fit to the high energy cutoff above 500 keV.}\label{fit_morepars2_fig}
\end{figure}

\begin{figure}
\plotone{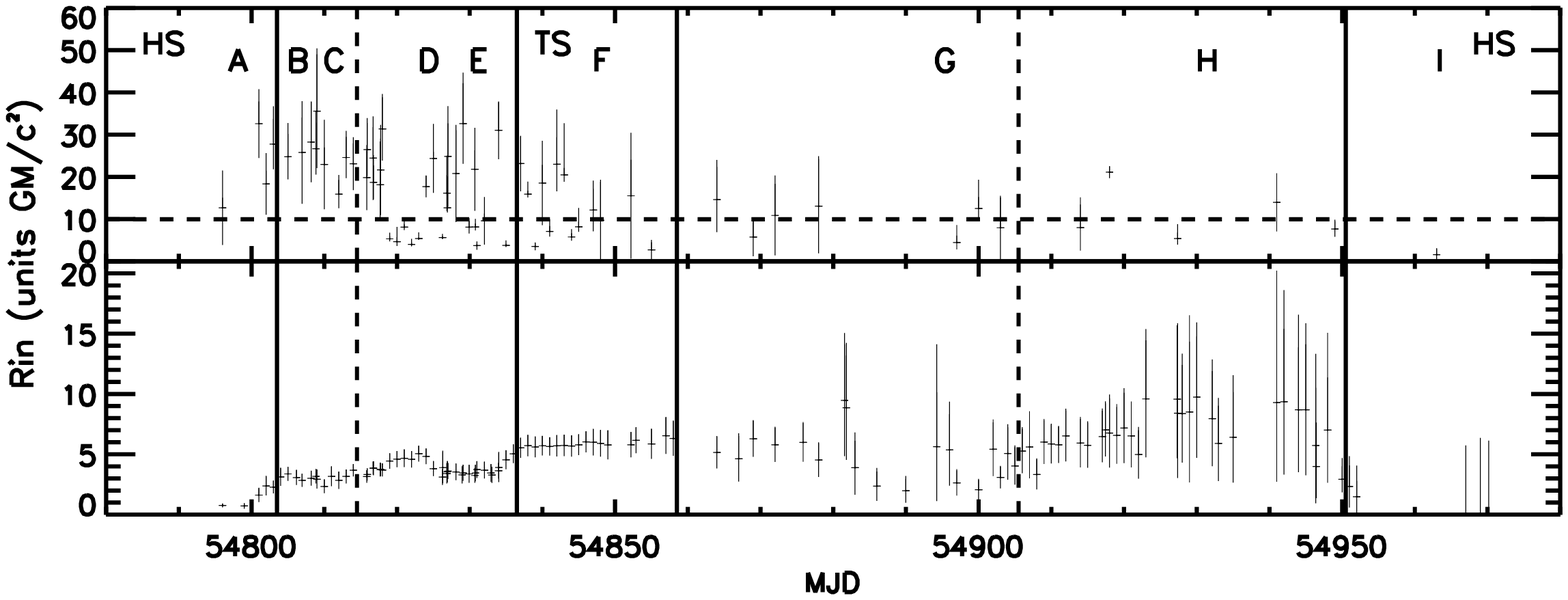} 
 \caption{Fits to the inner disk radius. {Top: $R_{in}$\ parameter from the {\em kdblur} model.  Bottom: $R_{in}$\ derived from the normalization to the {\em diskbb} model, assuming mass = $10 M_{\odot}$, $D = 10$~kpc, and $\cos(\theta) = 0.1$. The first two of these parameters are typical values for BH systems and the third one is based on our average fitted value.} 
The vertical lines and letters are as defined in the caption to Figure~\ref{lin_lc_fig}.  The horizontal line in the top plot shows the weighted mean.  }\label{fit_morepars1_fig}
\end{figure}

\begin{figure}
 \epsscale{1.1}
\plotone{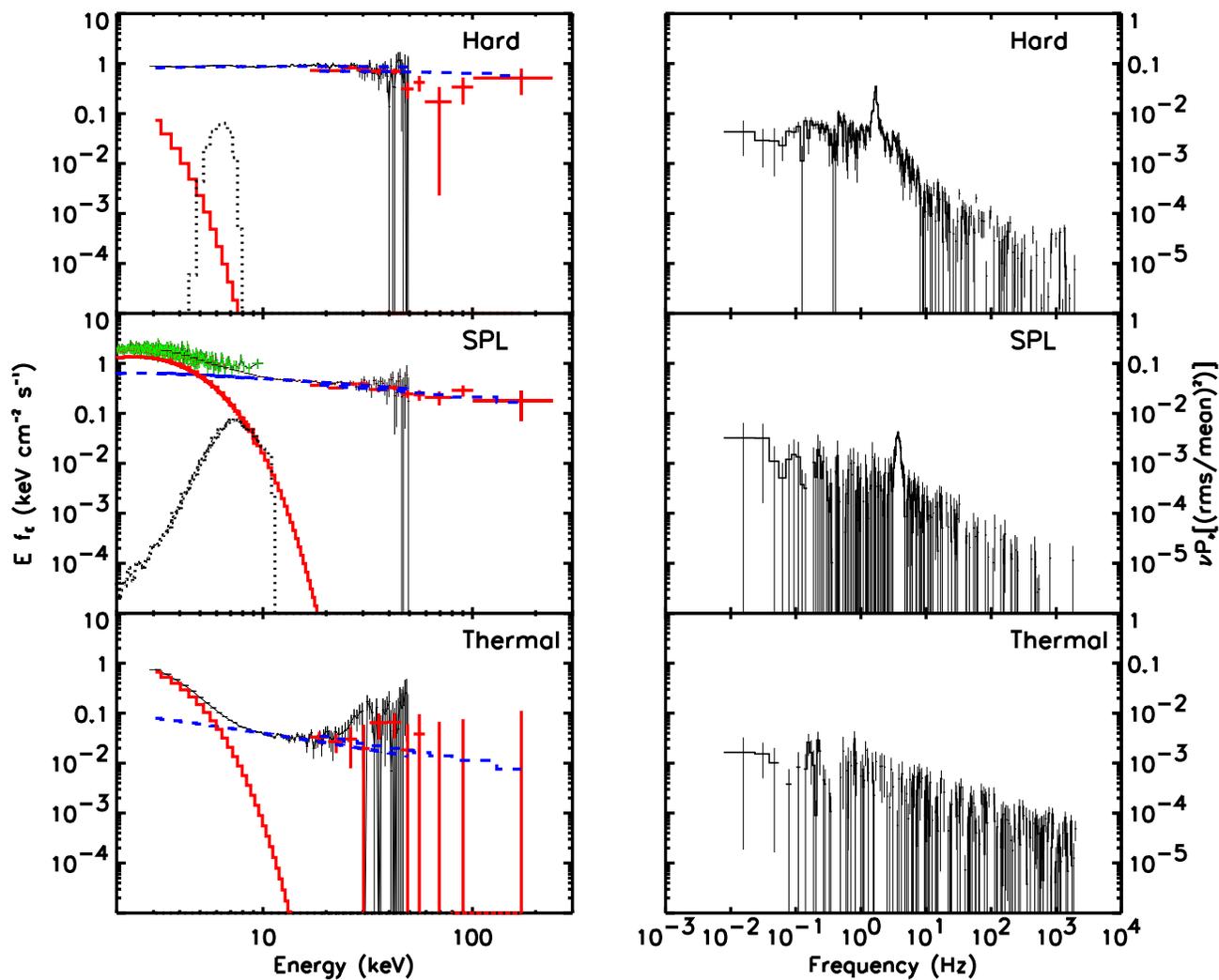} 
  \caption{Energy spectra (left) and power spectra (right) for three representative source states: hard (MJD 54801; 1-Dec-2008), SPL (MJD 54826; 26-Dec-2008) and thermal (MJD 54846; 15-Jan-2009). In the energy spectra, black points are PCA, red points are HEXTE and green points are XRT.  The component models are power-law (blue), black body (red) and iron line (black).  
  }\label{powerspec_fig}
\end{figure}

\noindent The average spectral parameters are shown in Table~\ref{tab-spect} and the well-constrained parameters are displayed in light curve format 
 in Figures~\ref{fit_pars_fig}, \ref{fit_morepars2_fig} and~\ref{fit_morepars1_fig}.   Representative energy spectra for three source states (see discussion in Section~\ref{discussion}) are shown in the left hand panels of Figure~\ref{powerspec_fig}.
 The second panel of Figure~\ref{fit_pars_fig} shows the PCA hardness defined as the ratio of the rate in the 5-12~keV band to the rate in the 3-5~keV band \citep[chosen to be consistent with the comparable plots in][]{mcre06}. This ratio varies between 0.2 and 1.5. It only rises above 1.0 during the first few days of the outburst, for a few days around MJD 54894, and then again at the very end of the observations.  During the time period from $\approx$\ MJD 54800 to 54875, the hardness ratio roughly follows the light curves in its decline with two rises at MJD 54830 (point {\em E} in Figure~\ref{fit_pars_fig}) and near MJD 54865, corresponding to rises in intensity in the BAT and HEXTE.  Around MJD 54894 ({\em G}) there is a significant hardening, matching again an increase in all of the light curves, before softening again. 

The correlation between hardness and intensity is shown in a different way in Figure~\ref{hr_fig}, where the X-ray hardness is defined slightly differently (see figure caption).  The trace recapitulates the main features of the second panel of Figure~\ref{fit_pars_fig} with a rapid softening early in the outburst, followed by subsequent hardening, once in the middle and once at the end of the outburst, all superimposed on a general decline in flux as the outburst decays.

The fit parameters for the two most significant and well-constrained model components are shown in the third and fourth panels of Figure~\ref{fit_pars_fig}.  The power-law index, $\Gamma$, starts low, before settling into a fairly constant range $2.1\ <\ \Gamma\ <\ 2.4$ with some excursions to lower values during the thermal state and higher values (though with poor statistics) late in the outburst, before dropping (hardening) again at the end of the outburst.  Our fit was to a power law with an exponential cutoff (Figure~\ref{fit_morepars2_fig}).  However, for most of the observations the cutoff energy was beyond the HEXTE energy range so it could not be fitted.  During the very early part of the outburst (before MJD 54820), we could fit a high-energy cutoff, and this parameter follows a pattern seen for other sources \citep[e.g.][]{mott09} where the cutoff energy falls gradually, here to 30~keV, with a sharp rise at the end of the hard state, followed by a flattening, here to $\approx 200$\ keV.  For mid-outburst observations it is not clear whether the cutoff energy is indeed $> 500$\ keV or whether low statistics in the HEXTE range prevent fitting this parameter.   There is some interesting behavior late in the outburst, when for some of the days after MJD  54930,  the data are consistent with a very low cutoff, averaging $7.2\ \pm  2.6$\ keV.  Though the result is of fairly low significance, it appears to be genuine and represents unusual behavior at this stage of a BH outburst.

For the entire outburst the disk temperature shows a steady cooling from $\approx 1$\ keV to 0.3 keV with some heating associated with rises in the PCA light curve between {\em C} and {\em D} and at {\em E} and {\em G}.

The bottom panel of Figure~\ref{fit_pars_fig} shows the relative fractions of the unabsorbed flux (2-20~keV) attributed to the two main models, the power law (shown in red) and disk (black).  This panel indicates that the power law dominates for most of the outburst, although it is above 75\% only during three episodes.  The disk fraction is above this level only for about 20 days around MJD 54850, and for long stretches the two components are of comparable intensity.

During all phases of the outburst, we see a feature which we interpret as an iron line at $6.72 \pm 0.58$\ keV.   The equivalent width of the line is consistently within the range expected for reflection (100 - 300 eV), which increases our confidence that we are using the right model. The fraction of the total flux in the line component is $< 5$\% and the normalization is $< 0.002\  {\rm photons}\ {\rm cm}^{-2} {\rm s}^{-1}$.  This feature is also seen in the XRT spectral analysis.

Results for the inner disk radius from the iron line (units $R_{in} \equiv GM/c^2$; Figure~\ref{fit_morepars1_fig}, top panel) show a lot of scatter and two clusters of values, high values near 25, particularly during the early hard part of the outburst, and low values near five.  Although not conclusive, this is consistent with the picture in which the disk inner radius is much larger in the hard state than in the softer states.  However, we cannot rule out that the bimodality may be a fitting artifact with two local minima in phase space.  {Using a faint iron line to infer the inner radius is susceptible  
to uncertainties related to changes in the illumination parameter as the X-ray state wanders.}

Another method sometimes used to derive $R_{in}$\ is to look at the normalization 
parameter of the {\em diskbb} model, $k\ =\ (R_{in}/D)^2\ \cos(\theta)$, where $R_{in}$\ is in km, $D$\ in units 10 kpc, and $\theta$\ is the inclination angle.  We have no information 
about $D$, so we cannot use this to directly determine $R_{in}$, but this analysis could tell us 
something about how $R_{in}$\ changes.  
The significant result from this is that  $R_{in}$\ appears constant during the thermal state (Figure~\ref{fit_morepars1_fig}, bottom panel), as expected.

At other times, this parameter shows what appear to be significant variations in a direction which contradicts the simple picture of inner disk contraction during the thermal state \citep[e.g.][]{esin97}.  However, one must be very careful in making such an interpretation.  Extracting a disk radius from {\em diskbb} is really only reliable in the thermal state.  Almost all the work done on finding BH parameters from the disk radius has concentrated on the thermal state.  Only very recently have attempts been made to extract disk radii from states other than the thermal state and the results have been mixed. In particular, Figure~1 in \citet{stein09a} shows that with any significant Comptonization (power-law) emission, the radius derivation becomes unreliable.  There are other difficulties as well.
 First, the disk black body could be modified by changing levels 
of electron scattering, which is adjusted using a ``color correction factor'' \citep{shim95,stein09}.
\citet{gier08} describe two other effects  
that can cause problems
with measuring  $R_{in}$\  using the thermal component: X-ray irradiation and  
the inner boundary
condition.
Finally, if the 
central black hole is spinning then one must use a relativistic analysis \citep[e.g.][]{stein10}. %
Such corrections 
are beyond the scope of this paper, so we are unable to draw any firm 
conclusions about the inner disk radius other than its general behavior during the thermal state.

\subsection{Timing}\label{results-timing}

\noindent All 160 power density spectra plots were scanned for evidence of periodic or quasi-periodic behavior.   Representative power spectra are shown in the right-hand panel of Figure~\ref{powerspec_fig}. There is no significant power above 10~Hz.  In Figure~\ref{power_fig} we show how the power density evolves with time.  The frequency of the QPO is shown in the top panel. Early in the outburst the frequency is quite low ($0.17\ \pm\  0.0005$\ Hz)
and rises rapidly over nine days to $6.3\ \pm  0.016$\ Hz.   There is quite a bit of scatter, but over the next 20~days, the QPO frequency remains between $\approx 4$\ and 8~Hz, before declining again.  For most of the period from MJD 54840 to 54890, the QPOs disappear before returning again with roughly the same pattern, a rise in frequency to $\geq 6$\ Hz, followed by another drop in frequency.  

The coherence parameter, $Q$ (second panel of Figure~\ref{power_fig}), was quite strong for most of the observations and remained near 20 for the early part of the outburst, decreasing after MJD 54815 to around 10.  As the outburst flux diminished, the QPOs weakened and it became more difficult to properly fit the peaks.  For most of the late observations (after MJD 54890) we could only fit the central frequency, not the width or coherence parameter. The total rms amplitude (power as a fraction of the mean count rate; bottom panel of Figure~\ref{power_fig}) starts quite high, remaining above 5\% for the first 16 days of the outburst, then is reduced somewhat, but stays at a few percent until the QPOs disappear.

The continuum power was also quite high during the early times, ranging between 10 and 30\% before dropping quite abruptly at MJD 54814 to an average of 4\% before returning to a higher level around MJD 54880 and then dropping below 10\% again at the HIMS-SIMS transition (see Section~\ref{discussion-temporal}).  Late in the outburst, the rms power remained low even though other indicators (power-law index, disk fraction) indicated a return to the hard state.

\section{Discussion}\label{discussion}

\subsection{State Transitions}\label{discussion-states}

\subsubsection{Temporal Evolution}\label{discussion-temporal}

\begin{figure}
\epsscale{1.1}
\plotone{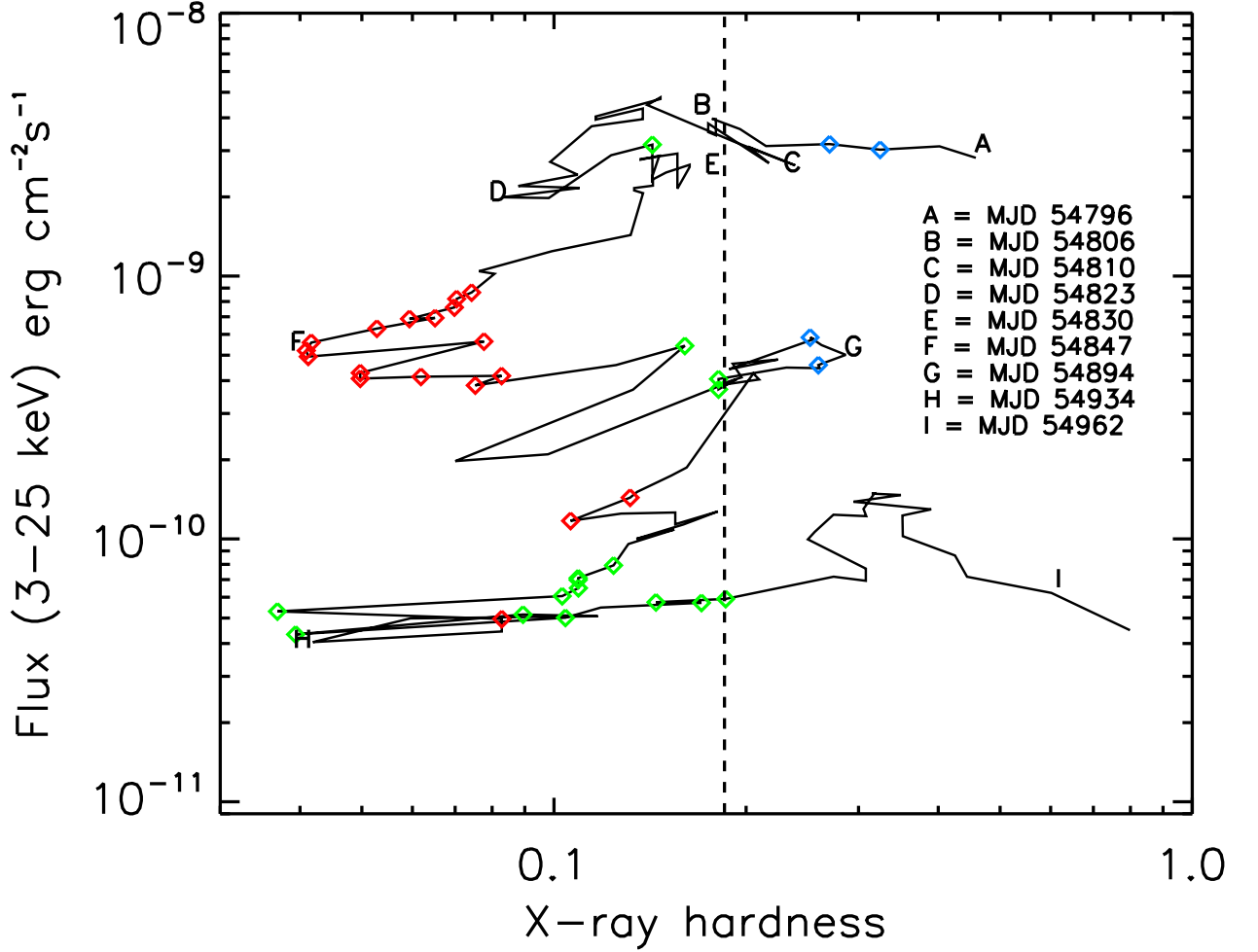}  
  \caption{Hardness-intensity diagram for Swift J1539.2$-$6227 showing the flux in the 3-25 keV band versus the X-ray color (9.14--25.11 keV/2.87--9.14 keV).  The letters represent major changes in direction of the trace.  The dashed line indicates the transitions between the hard-intermediate (to the right) and soft-intermediate states (to the left).   Red points indicate the thermal state, green SPL, and blue the hard state as derived using the definitions of \citeauthor*{remc06} (see Section~\ref{discussion-states}). Note that these definitions do not always correspond to more traditional definitions of spectral states.}\label{hr_fig}
\end{figure}

\begin{figure}
\epsscale{1.1}
\plotone{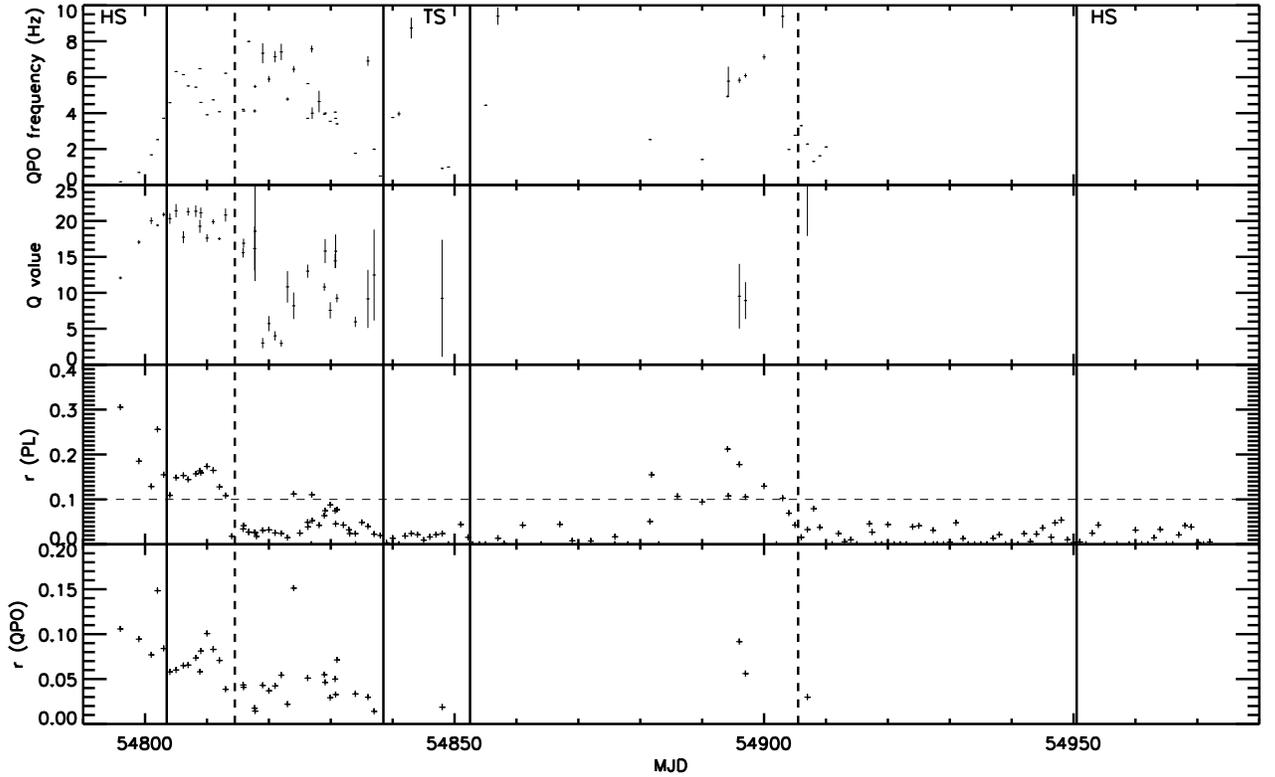} 
  \caption{Results of timing analysis.  The power continuum is integrated from 0.1 to 10 Hz. Strong QPOs continue even into the possible thermally dominated regime, then disappear around MJD 54850, reappearing when the source again hardens.  Although we expect a return of a QPO signal late in the outburst (after MJD 54950), the overall count rate is too low for a QPO peak to be detected.  The vertical lines are the same as in Figure~\ref{lin_lc_fig}.
  }\label{power_fig}
\end{figure}

\noindent Swift J1539.2$-$6227 appears to go through a fairly typical sequence of transitions during its outburst.  Because they are more physically meaningful and also because the common ``low'' and ``high'' appellations are not applicable to X-ray rates measured above $\sim 10$\ keV, we  adopt the naming conventions of \citeauthor*{remc06}: the hard state (a.k.a. low-hard) which is characterized by a hard power-law (PL) component and QPOs and in which the accretion disk is faint and cool; the thermal state (high-soft) which is  dominated by heat radiation from the inner accretion disk; and the steep power-law (SPL) state (very high) which has significant  contributions from both PL and thermal emission and is often accompanied by QPOs. {We also apply the state definitions of \citeauthor*[][Table 2]{remc06}, to determine when Swift J1539.2$-$6227 entered and left the canonical BH transient states and when it was in some other intermediate state.}
We note the signature of a hard state both early and late in the outburst{, a thermal state in the middle of the outburst and, in between, relatively rapid transitions back and forth between harder and softer intermediate states.}
We note that if we apply the strict definitions of \citeauthor*{remc06}, most of the time Swift J1539.2$-$6227 is in an intermediate state -- often one or two of the criteria are met, but another (most often the power-law index) is not.  To further illustrate the state transitions we derive a hardness-intensity diagram (HID; Figure~\ref{hr_fig}) similar to those in \citet{fend04} and \citet{mott09}.  The letters on the plot, and in other figures in the paper, correspond to significant turning points in the trace. 

\noindent {\bf Initial Hard 
State.} The source is first seen to be rapidly brightening in  what is likely a short-lived hard state.  This state corresponds to the highest peak in both BAT (15-50 keV) and HEXTE (16-100 keV) and is indicated in Figures~\ref{lin_lc_fig} -- \ref{fit_morepars1_fig} and \ref{power_fig} as the time before the first vertical line. 

The hardness ratio HR (5-12 keV)/(3-5 keV) starts near 1.5, which is the value \citet{mcre06} give as characteristic of the hard state.  By the end of the rise, HR has dropped to 0.7, with a upturn near {\em C}, seen most clearly in Figure~\ref{hr_fig}.  The photon power-law index also shows a trend from hard (1.1) to soft (2.2) during this period.  {The power-law cutoff energy first falls slowly and then rises rapidly from $\approx 40$\ keV to around 200~keV.}
The energy spectrum is dominated by the PL component ($> 80$\% of the flux) and the disk temperature shows a slight drop from 1.2~keV to 0.9~keV.   This is also the period during which the QPO frequency increases rapidly from 0.17~Hz to 6.3~Hz and the QPO peaks sharpen.  
{The QPOs are strong ($r > 5$\%) as is the power spectrum continuum ($r > 10$\%).
Note that in Figure~\ref{hr_fig} we do not see the long vertical segment which would precede point {\em A} and correspond to the canonical early hard state measured for many other sources.   Therefore, this diagram suggests that the transition in Swift J1539.2$-$6227 begins at point {\em A} and is largely complete by point {\em B} and that we do not observe (other than in the BAT) the rise to the peak of the hard state.  
Many aspects of a hard state persist until point {\em C}, showing }
signs of being what \citet{bell10} would call a hard intermediate state.  

We can compare {the hard state} in Swift J1539.2$-$6227 with the evolution of  {spectral} parameters in the 2006-2007 outburst of the black hole candidate GX 339$-$4, as discussed by 
 \citet{mott09}, though with different state names and definitions.  One can see very similar trends between the two sources. 
 GX 339$-$4 shows a full transition {through the} hard state over about eight days, with the hardness ratio falling and power-law index rising steadily during the transition.  
The cut-off energy first drops slowly then rises sharply.
All of these parameters then flatten out during the hard (high-soft) state.  We see quite similar behavior of the three parameters for Swift J1539.2$-$6227 during the initial hard state 
over a similar timescale, $\approx 5$\ days.  
One very striking difference between the sources is that the initial rise in the BAT count rate for GX 339$-$4 preceded the hard-soft transition by $\approx 80$\ days, whereas in Swift J1539.2$-$6227, the rise is  $\approx 1$\ day. This is not simply a threshold effect since the peak BAT count rate of the 2006-2007 GX 339$-$4 outburst is only about twice that of the peak count rate for Swift J1539.2$-$6227.  We note that GX 339$-$4 is a persistent source with repeated outbursts which may well have different pre-outburst behavior in the hard X rays.
 
\noindent {\bf Themal State.} Around MJD 54835, several parameters (though not all) change sharply.  The QPOs disappear, the spectral index falls and the disk fraction begins to dominate, rising above 75\%.  There is no sharp change in either the temperature or integrated power continuum.  The HR reaches its lowest value of $\approx 0.2$.   This is clearly now a thermal state which lasts for about 20 days.

\noindent {\bf Intermediate States.} {Swift J1539.2$-$6227 is similar to other sources in that the time between clearly defined states is marked by a series of intermediate states, where the HR and other parameters move back and forth in phase space.  We will examine a few of the most significant such episodes.}

At MJD 54814 (dashed vertical line in Figures~\ref{lin_lc_fig} -- \ref{fit_morepars1_fig} and \ref{hr_fig}) there is an abrupt drop in the continuum power and a broadening of the QPOs, indicating a shift from Type C to Type B QPOs \citep[see][for the definition]{case05}.  This change in timing properties, unaccompanied by major changes in energy spectral parameters, is what \citet{bell10} defines as the transition from a Hard-Intermediate (HIMS) to Soft-Intermediate state (SIMS).  We note that there is a change in the relative proportions of the disk and PL contributions a few days later with the PL clearly dominant before and the two components nearly equal after.  
It is possible that the source approaches or even reaches the SPL state at least once during this period, near {\em E}, where the PL index rises to $\Gamma\ =\ 2.4$.

After roughly MJD 54878 (around {\em G}), there is an increase in { count rate,}  PL fraction and hardness ratios.  
There is also a significant increase in the power continuum followed about ten days later by a return of the low-frequency QPOs, although they appear less significant now due to the factor of $\sim10$\ reduction in rates.    However, comparing this period (around {\em G}) to the period just before the first vertical line, one sees that while the source conditions bear some similarity to a hard state, neither the hardness ratio nor the PL index are fully indicative of a return to a hard state.  Thus we conclude that this is a temporary hardening, similar to what is seen for XTE J1859+226 \citep{broc02}, in which short-lived spectral hardening was accompanied by radio ejections.  Then at around MJD 54904, there is another abrupt drop in continuum power which suggests a further crossing of the HIMS-SIMS ``jet line.''

\noindent {\bf Final hard state.} After about MJD 54945, there is {another} sharp drop in the power-law index to $< 2.0$\ along with a continued decline in disk temperature.  The hardness ratio also starts to rise above 1.0 at this time and there is a peak in both the PCA and BAT light curves at around MJD 54950.  There is {some} indication that there is again a low-energy spectral energy cut-off after MJD 54930, though the statistics are poor.  The change in PL index and HR, together with the trace {\em H-I} in the HID suggest that the source is making a transition back to the hard state.   

{However,} the behavior of the power continuum level late in the outburst engenders some doubt as to whether we have indeed observed a return to the hard state.  
{No break was found in the power spectra at this stage, but it was still possible to integrate the continuum and we find that, as seen in Figure~\ref{power_fig}, the continuum power level remains low ($< 10$\%).}
{Furthermore} the PCA count rate is so low {at this time} that one cannot reliably derive a hardness ratio, so the {HID} trace  {(Fig~\ref{hr_fig})} ends before reaching the vertical segment seen in, for example, \citet{mott09}.  In short, although it is quite likely that Swift J1539.2$-$6227 has reached the hard state by MJD 54950, the possibility remains that we are seeing only a temporary hardening (like near {\em G} in the HID) or a continued intermediate state on its way to the hard state.

By the time the PCA observations ended, the key parameters (hardness ratio and spectral index) had not yet settled out.  Fits to the spectra of the final {outburst} observation at MJD 55038 showed the index and disk temperature in the same range as before the data gap. We note that the PCA and HEXTE fluxes are nearly the same at this late point as they were before, even after the subtraction of the background rate, suggesting that the source itself may still be active. However, the BAT light curve shows no detection over most of the period with no PCA observations and only a weak detection at MJD 55038.
 
 \subsubsection{Parameter Correlations}

{In Figure~\ref{randm_fig}} we  plot the three most significant parameters for determining state: power-law index, disk fraction and continuum power, versus X-ray hardness and indicate therein {and in Figure~\ref{hr_fig}} the observations meeting the criteria for each of the three states as defined in \citeauthor*{remc06}.

There are some inconsistencies between the \citeauthor*{remc06} definitions and other typical state definitions.  In Figure~\ref{hr_fig}, hard state points are found only at two parts of the outburst.  For the two earliest observations, we find $\Gamma\ <\ 1.4$, which is below the minimum set by  \citeauthor*{remc06}, although by most other criteria this time would certainly be considered a hard state.  Then there appears to be a return to a hard state near {\em G}, where the PL index dips below 2.1, coincident with a rise in PL fraction and power continuum level.  An actual return to a hard state in the middle of an outburst would be unprecedented, and we do not claim, based on just this set of parameters that this has occurred.  Rather we think that natural scatter in the fit parameters as the spectrum hardens has pushed a few points into the hard regime.  Also, at the very end of the outburst, the formal definition is not met as the power law continuum level remains below 10\%.

Likewise the forays into short SPL states should be viewed with caution, especially those with low values of X-ray hardness.  There are scattered episodes both before and after the thermal state and an extended period near {\em H}.  However, even at this point, there are large errors on the PL index and timing features are too weak to be meaningful for classification.  So it is fair to say that it is likely, though not conclusive, that Swift J1539.2$-$6227 had brief episodes of the SPL state.  The thermal state near {\em F} appears unambiguous and all criteria converge on the same classification.  The few other thermal state points (around {\em H}) again could well be fluctuations at times of low statistics.  It is worth noting that several of the sources profiled in \citeauthor*{remc06} had similar episodes of apparent multiple brief state changes.  Finally, this classification system leaves many of the observations unclassified and in some intermediate state, a situation found by other authors for other sources \citep[e.g.][]{mott09}.

Despite the limitations  of these state definitions, it is useful to compare Swift J1539.2$-$6227 to the six other sources studied in \citeauthor*{remc06} using the three plots in Figure~\ref{randm_fig}.  We first examine PL index versus hardness.  We do not see the very steep index in the thermal state that \citeauthor*{remc06} show for most of the sources and most thermal state points fall under the SPL points, as seen for XTE~J1550$-$564, H~1743$-$322 and XTE~J1859+226.  In fact our plot most closely resembles that for XTE~J1859+226 with a scattering of thermal points to the left with PL index $\approx 2$, some SPL points above and to the right followed by a large number of intermediate state points with a scattering of hard points to the far right. The disk fraction plot shows a single trace which nicely matches the patterns shown for those sources in  \citeauthor*{remc06} with simple behavior, such as GRO~J1655$-$40 and 4U~1543$-$47.  The rms-hardness plot is less easy to interpret since many of the SPL and hard state points do not have enough flux to constrain the rms values.  Nonetheless, the overall pattern is seen in which thermal points cluster to the lower left with rising rms through SPL to hard state as in GX~339$-$4.

We see the same general pattern here as for all six  \citeauthor*{remc06} sources, namely strong correlation of disk fraction and rms with hardness and separations in the PL-hardness space for the various states, showing clearly that Swift~J1539.2$-$6227 has the hallmarks of a BH transient. However, we cannot use the plots to conclusively compare this outburst to any particular earlier outburst of another source.

\subsection{Light Curve Correlations}\label{discussion-correl}

\noindent  Examination of Figure~\ref{lin_lc_fig} shows that the flux peaked in the BAT and HEXTE energy ranges before it peaked in the PCA and ASM.  We have attempted to determine the temporal shift in the light curves between the various energy bands.  First of all, to determine the delay during the first hard state and the following intermediate state, we shifted the BAT light curve forward in time by $\Delta T$\ and correlated the shifted light curve with the PCA light curve.  We determined that the correlation function peaked at $\rho = 0.82$ (Spearman rank correlation coefficient) for $\Delta T$\  = 8.0 d.  The correlation is shown in Figure~\ref{corr2_fig}, solid points.  This shift is quite similar to what was found for GRO J1655$-$40 \citep[7 d;][]{broc06}. 
After the transition to the soft intermediate state, the light curves tracked each other more closely.  Using $\Delta T$\ = -1.0 d (BAT shifted to {\em earlier} times), we find $\rho = 0.84$\  for the soft intermediate state (open squares in Figure~\ref{corr2_fig}).  Then during the thermal state, the correlation is somewhat less ($\rho = 0.71$; open diamonds) as the BAT count rate flattens out.  Taking the whole SIM plus thermal state, one gets $\rho = 0.92$.  This is in contrast to the outburst of GRO J1655$-$40; \citep{broc06}, in which the high and low energy light curves do not track each other during the thermal state.  In GRO~J1655$-$40 the difference between the high and low energy X-ray light curves is related to the relative contributions from the jet/corona and disk components. There seemed to be a significant contribution to the disk emission at XRT and ASM energies, hence the hard and soft X-ray light curves did not track each other. When the hard and soft light curves do track each other, as in Swift~J1539.2$-$6227, then it suggests  (although correlation does not necessarily imply a causal link) that the non-thermal (i.e. corona/jet) emission is dominating the soft X rays as well as the hard X rays -- the temperature of the disk emission is too low for this component to dominate at BAT energies.  This is consistent with what is seen with the disk and power law contributions during the intermediate states (bottom panel, Figure~\ref{fit_pars_fig}).  So Swift~J1539.2$-$6227 differs from GRO~J1655$-$40 in that either the disk emission is weaker or it peaks at lower energies, following the ``outside-in" disk outburst mechanism \citep[][discussion section]{broc06}.

	\section{Summary}\label{summary}

\noindent We present a complete progression of spectral and timing property evolution during the entire 2008-2009 discovery outburst of Swift J1539.2$-$6227.  Although a definitive classification of this source as a black hole binary transient awaits an estimate of the mass of the compact object, the many similarities between the outburst states and transitions in Swift J1539.2$-$6227 and those of other BH sources makes this new source a strong black hole candidate.   {We note also that typical a neutron star hardness-intensity diagram \citep[e.g.][]{vand06} shows very different behavior.  Z sources trace out tracks that are roughly Z-shaped and on relatively short (hour to day) timescales.  Atoll sources move more slowly than Z sources (timescales of days to weeks) and some sources such as Aql X-1 show a circular shaped pattern with hysteresis, similar to a BH source, though lacking the hard, high luminosity region.  Atoll sources also typically have type I X-ray bursts, none of which were observed in Swift J1539.2$-$6227. In general, the progression of state transitions in neutron stars is more rapid than what is observed in Swift J1539.2$-$6227, adding further support to its BH identification.}

\begin{figure}
\epsscale{0.75}
\plotone{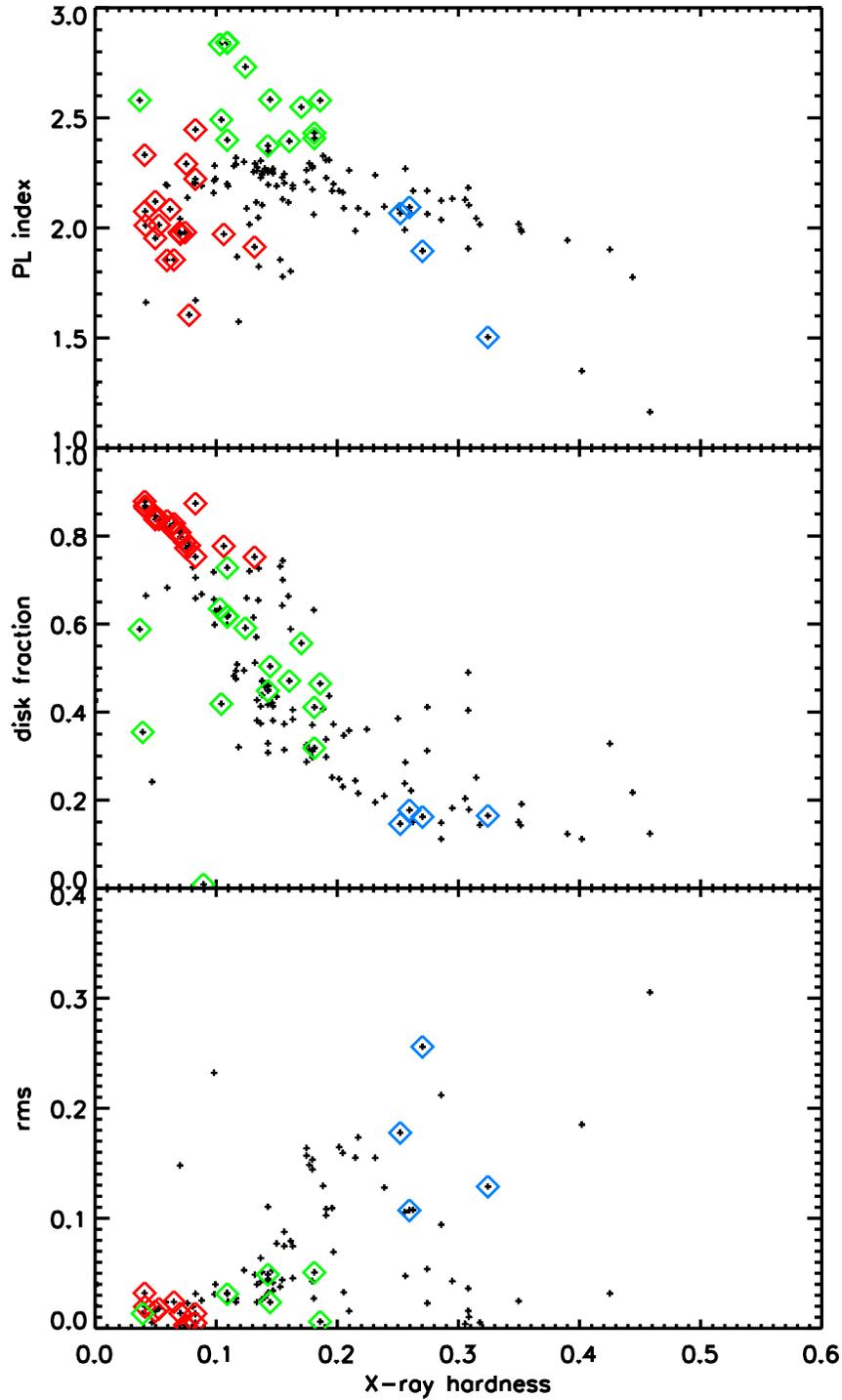} 
  \caption{Three prominent outburst properties are plotted with respect to X-ray hardness (9.14--25.11 keV/2.87--9.14 keV).  The colors indicate the source state determined using the definitions of \citeauthor*{remc06}.  Red points indicate the thermal state, green SPL and blue the hard state. %
    }\label{randm_fig}
\end{figure}

\begin{figure}
\plotone{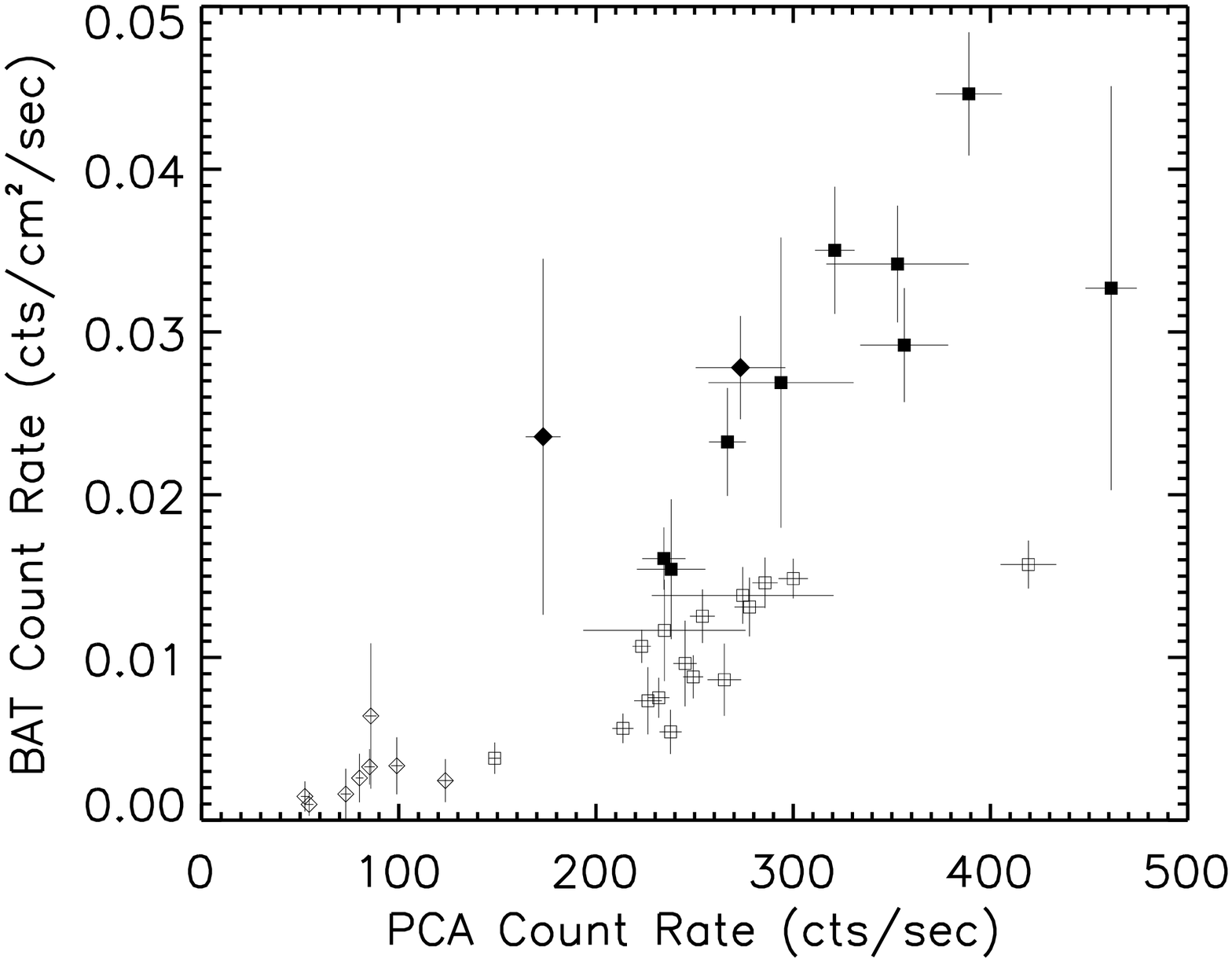} 
 \caption{The correlation between the BAT count rate and the PCA count rate for four episodes during the outburst.  The solid points are from MJD 54796 -- 54815 (the hard state as diamonds and the hard intermediate state as squares) with the BAT rates shifted forward by 8.5 days.  The open points are from MJD  54815 -- 54859 (the soft intermediate state as squares and thermal state as diamonds) with the BAT rates shifted backward by 1.0 day.
  }\label{corr2_fig}
\end{figure}

Although the details vary depending on the physical properties (such as mass accretion rate) of the individual binary system begin studied, observations of BH and BH candidate outbursts reveal many commonalities.  The Swift J1539.2$-$6227 outburst can be compared to a number of other cases.  The rapid rise of the hard X rays ($\approx 1$\ day) and the hard flux preceding the soft flux by $\approx 8$\ days is similar to what has been observed for GRO~J1655$-$40 \citep{harm95,broc06}, for XTE~J1859+266 \citep{broc02,fend04} and for other sources.  In both of these named sources and in XTE~J1550$-$564 \citep{fend04}, the hard X rays preceded the soft X rays by a few days; similar behavior was seen in Swift J1539.2$-$6227.    Another similar observed phenomenon is the presence of periods of temporary hardening in the middle of the outburst, seen clearly in  XTE~J1859+266, GRO~J1655$-$40 and now Swift J1539.2$-$6227.   Also comparison of Figure~\ref{randm_fig} with the corresponding plots in \citeauthor*{remc06} shows clear similarities between this new source and other well-studied BH systems in the correlation of power-law index, disk fraction and rms power with X-ray hardness.

Not only does the Swift J1539.2$-$6227 outburst share these broad features with other sources, one can also see a number of similarities on finer scales in tracking the evolution of individual spectral and timing properties, in particular during the transition from the hard state to the hard intermediate state.  Comparing Swift J1539.2$-$6227 to GX 339$-$4 \citep{mott09} and H~1743$-$322 \citep{prat09}, we see a rapid increase in photon index, and a gradual fall and then rapid rise in cutoff energy.   Although the statistics on the fit to the inner disk radius are not good, there is an indication that the disk radius decreases from $\approx 25$\ in the hard state to $\approx 5$\  in the thermal state (units $GM/c^2$) and remains constant in the thermal state, all consistent with disk models \citep[e.g.][]{mcre06}.
With the timing properties the correspondence is quite good between Swift J1539.2$-$6227 and H~1743$-$322 \citep{prat09}.  The QPO frequency rises to a fairly steady level at $\approx 6$\ Hz, while the total power and QPO power both decrease over the transition.  Also the $\approx 6$\ Hz QPO frequency in the intermediate state appears to be a common BH feature which \citet{case05} conclude is related to some fundamental process.   Later on, a clear step down in total rms power without a change in QPO frequency in Swift J1539.2$-$6227 indicates a HIMS to SIMS transition. 

All of these similarities, combined with the lack of observed pulsations, give strong weight to the conclusion that Swift J1539.2$-$6227 is a BH candidate.  The interpretation of the states is consistent with previous work.  Although we do not have the radio observations to prove the existence of jets, we see other signs of an initial hard state accompanied by low-frequency QPOs, a state which likely arises from an active corona component producing X rays either through synchrotron or Compton emission (or likely both) \citep[e.g. RM06;][]{tita04,broc06,mott09,fend04}. The outburst also spent $\approx 20$\ days in a thermal state arising from thermal emission from a slowly brightening and then fading accretion disk \citep[RM06;][]{broc06}.  It is an instability in this disk which is believed to be the trigger for the entire outburst (\citeauthor*{remc06}).  Although it is not absolutely clear whether Swift J1539.2$-$6227 entered the SPL state during its outburst, particularly given the weakness of the power spectrum at these times, there is some evidence that this BH outburst state was reached as well.

\acknowledgments 
H.A.K. and C.B.M. are supported by the {\em Swift} project.  J.A.T. acknowledges partial support from NASA under {\em Swift}  
Guest Observer grant NNX08AW35G.  The authors gratefully acknowledge the {\em RXTE} and {\em Swift} principal investigators for approving, and mission planners for scheduling, the many observations used for this work.


\begin{deluxetable}{llc}
\tablewidth{0pt} 	      	
\tabletypesize{\scriptsize} 
\tablecaption{{\em Swift} pointed observations of Swift J1539.2$-$6227\label{tab-xrt}}
\tablehead{\colhead{Date} & \colhead{MJD} & \colhead{Exposure (s)}}
\startdata
2008-Dec-25 & 54825 & 3944\\
2008-Dec-26 & 54826 & 3320\\
2009-Mar-08 & 54898 & 2490\\
2009-Mar-15 & 54905 & 1683\\
2009-Apr-11 & 54932 & 4831\\
2009-May-15 & 54966 & 2567\\
\enddata
\end{deluxetable}

\begin{deluxetable}{llclcc}
\tablewidth{0pt} 	      	
\tabletypesize{\scriptsize} 
\tablecaption{Average spectral fits for Swift J1539.2$-$6227\label{tab-spect}}
\tablehead{\colhead{Model} & \colhead{Parameter} & \colhead{Units} & \colhead{Mean} & \colhead{Spectra used\tablenotemark{a} } & \colhead{Notes}}
\startdata
wabs & $N_H$ & atoms/${\rm cm}^2$ & $3.5 \times 10^{21}$ & -- & frozen \\
diskbb & Inner disk temperature & keV & -- & 146 & see Figure~\ref{fit_pars_fig} \\
diskbb & normalization & -- & -- & 146 & see Figure~\ref{fit_morepars1_fig} \\
cutoffpl & Photon index & -- & -- & 148 & see Figure~\ref{fit_pars_fig} \\
cutoffpl & High energy cut-off & keV & -- & 20 & see Figure~\ref{fit_morepars2_fig} \\
kdblur & Emissivity index & -- & $2.84 \pm 0.56$ & 61 &  {frozen in final run}\\
kdblur & Inclination & degrees & $82 \pm 8$ & 116 &  {frozen in final run}\\
kdblur & Inner radius & $(GM/c^2)$ & $9.9 \pm 7.8$ & 66 & see Figure~\ref{fit_morepars1_fig} \\
kdblur & Outer radius & $(GM/c^2)$ & 100.0 & -- & frozen \\
gaussian & Line energy & keV & $6.71 \pm 0.65$ & 103 & \\
gaussian & Line width & keV & 0.0 & -- & frozen \\
\enddata
\tablenotetext{a}{This column gives the number of spectra (out of 150 total) in which the particular parameter was reasonably well constrained.} 
\end{deluxetable}

\clearpage
\epsscale{0.9}

\clearpage

\end{document}